\documentclass[conference]{IEEEtran}
\IEEEoverridecommandlockouts
\usepackage{cite}
\usepackage{amsmath,amssymb,amsfonts}
\usepackage{dirtytalk}
\usepackage{multirow}
\usepackage{algorithmic}
\usepackage{graphicx, graphics}
\usepackage{textcomp}
\usepackage{xcolor}
\usepackage{hyperref}
\def\BibTeX{{\rm B\kern-.05em{\sc i\kern-.025em b}\kern-.08em
    T\kern-.1667em\lower.7ex\hbox{E}\kern-.125emX}}

\DeclareMathOperator*{\argmin}{arg\,min}

\makeatletter
\def\@IEEEpubidpullup{8\baselineskip}
\makeatother
\begin{document}

\IEEEoverridecommandlockouts
\IEEEpubid{
\parbox{\columnwidth}{\vspace{-6\baselineskip}Permission to make digital or hard copies of all or part of this work for personal or classroom use is granted without fee provided that copies are not made or distributed for profit or commercial advantage and that copies bear this notice and the full citation on the first page. Copyrights for components of this work owned by others than ACM must be honored. Abstracting with credit is permitted. To copy otherwise, or republish, to post on servers or to redistribute to lists, requires prior specific permission and/or a fee. Request permissions from \href{mailto:permissions@acm.org}{permissions@acm.org}.\hfill\vspace{0.8\baselineskip}\\
% \begin{spacing}{1.2}
{
\small\textit{ASONAM '19}, August 27-30, 2019, Vancouver, Canada \\
\copyright\space 2019 Association for Computing Machinery. \\
ACM ISBN 978-1-4503-6868-1/19/08 \$15.00\\
\url{http://dx.doi.org/10.1145/3341161.3343686}
% \end{spacing}
}
\hfill}
\hspace{0.9\columnsep}\makebox[\columnwidth]{\hfill}}
\IEEEpubidadjcol

\title{On the Structural Properties of Social Networks and their Measurement-calibrated Synthetic Counterparts}

\author{\IEEEauthorblockN{Marcell Nagy and Roland Molontay}

\IEEEauthorblockA{MTA-BME Stochastics Research Group, Budapest, Hungary
}
\IEEEauthorblockA{Department of Stochastics, Budapest University of Technology and Economics, Hungary}
\IEEEauthorblockA{Faculty of Informatics, University of Debrecen, Hungary
}
\IEEEauthorblockA{Email: marcessz@math.bme.hu, molontay@math.bme.hu}}

\maketitle

\begin{abstract}
Data-driven analysis of large social networks has attracted a great deal of research interest. In this paper, we investigate 120 real social networks and their measurement-calibrated synthetic counterparts generated by four well-known network models. We investigate the structural properties of the networks revealing the correlation profiles of graph metrics across various social domains (friendship networks, communication networks, and collaboration networks). We find that the correlation patterns differ across domains. We identify a non-redundant set of metrics to describe social networks. We study which topological characteristics of real networks the models can or cannot capture. We find that the goodness-of-fit of the network models depends on the domains. Furthermore, while 2K and stochastic block models lack the capability of generating graphs with large diameter and high clustering coefficient at the same time, they can still be used to  mimic social networks relatively efficiently. 
\end{abstract}


\section{Introduction}
\label{sec:intro}

The study of complex networks started with social network analysis. Many of the most important topological network characteristics have their roots in the social domain. For example, the small-world phenomenon often popularized as \say{six degrees of separation} was originally set out  for the network of personal acquaintances by the Hungarian novelist, Frigyes Karinthy in 1929~\cite{karinthy1929chain}. Later Karinthy's intuition was confirmed by Milgram's famous experiment as well~\cite{milgram1967small}. High clustering coefficient can also be clearly demonstrated by a social network example: \say{the friends of my friends are also likely to be my friends}. Some of the most influential network models were also motivated by social phenomena such as Watts--Strogatz small-world model~\cite{watts1998collective}, Barab\'asi--Albert preferential attachment model~\cite{barabasi1999emergence}, the stochastic block model~\cite{holland1983stochastic}, and the caveman graph~\cite{watts2004small}. Such models are mathematically tractable, moreover, an appropriate model captures the key characteristics, yet preserves the privacy of the original social network (e.g. Facebook friendship network) \cite{fong2009privacy}.  

Increasing computing power has made it possible to analyze massive datasets of large social networks using data-driven approaches. This paper also follows this line of research by analyzing 120 large social networks from three domains: friendship networks and communication networks mainly from social media sites, moreover collaboration networks (mostly scientific co-authorship graphs). We analyze the structural properties of these real-world networks, with an emphasis on revealing the correlation structure of the graph metrics across domains. 

After identifying a small selection of metrics that describe the networks well enough, we generate four synthetic graphs for each real-work network using four well-known network models (clustering Barab\'asi--Albert~\cite{holme2002growing}, stochastic block model \cite{peixoto2017nonparametric}, forest-fire~\cite{leskovec2005graphs}, 2K~\cite{gjoka2015construction}) with measurement-calibrated parameters to ensure that the model-generated counterparts are as similar to the real-world networks as possible -- according to the previously selected graph metrics. We study the goodness-of-fit of the models and identify the structural relations of real social networks that models can or cannot capture. The present study is an extension of an earlier paper of the authors~\cite{nagy2018data}. Using similar techniques, here we solely focus on the structural analysis of real and model-generated networks representing social structures thus both the real networks and network models studied in this work differ from the ones considered in~\cite{nagy2018data}.

Data-driven analysis of social networks has received a lot of research interest recently, mainly due to the growth of social network sites that revolutionized the availability and amount of social data. Note that a large amount of related literature focusing on online social networks carry out investigations from a sociological point of view, while here we focus on the structural properties from a network theoretical perspective. 

Traud \textit{et al.}~\cite{traud2012social} give a brief review of related sociological works and investigate the social structure and properties of Facebook friendship networks of 100 American universities, where various attributes are associated with the nodes such as gender, class year, major, etc. Sala \textit{et al.}~\cite{sala2010measurement} apply a grid search based measurement calibration technique to fit six network models to four large Facebook networks.  

Janssen \textit{et al.}~\cite{janssen2012model} and Bl\"asius \textit{et al.}~\cite{blasius2018towards}  introduce machine learning based techniques to select a network model that best fits a target real network and they also evaluate their method on Facebook networks and find that 2K model can efficiently mimic social networks~\cite{janssen2012model}, while  Chung-Lu and hyperbolic random graph models are bad representations of Facebook networks~\cite{blasius2018towards}. Faust in~\cite{faust2006comparing} studies and compares 51 social networks using a motif counting based approach~\cite{faust2006comparing}. There are also numerous papers concerned with network embedding, for instance in a recent work Liao \textit{et al.}~\cite{liao2018attributed} introduce a novel attributed network embedding method for social networks which utilizes the unique characteristics of social networks such as the network homophily phenomenon. 

As the aforecited papers show that in recent years, data-driven analysis of social networks has received extensive research interest from various perspectives. A key contribution of this work is that we propose a new methodology that unifies more branches of data-based social network analysis, such as the study of the relationship of metrics, network similarity, and model calibration. The fact that this large-scale study relies on the analysis of 120 real-world networks and 480 model-generated synthetic graphs also enables us to explore the correlation profile of the structural properties.

\section{Data and methodology}
\label{sec:data_method}

This study is based on 17 structural measurements of 120 large social networks, that are available in the supplementary material together with a short description~\cite{own_data2}.  The considered 17 graph measurements are chosen in such a way that together they measure every aspect of networks, such as degree distribution related metrics (e.g. density and interval degree probabilities), shortest paths (e.g. diameter), centralities (e.g. maximum eigenvector centrality), and clustering metrics (e.g. global clustering coeff.). Note that the considered metrics are in alignment with related studies~\cite{bounova2012overview, garcia2013correlation}. Table~\ref{table:networks} gives a brief overview of the collected data that were gathered from online network repositories~\cite{networkrepository, icon, konect}. Here the networks are considered as simple, undirected, unweighted graphs. 

\begin{table}[ht]
\caption{Composition of the collected set of real networks.}
\scalebox{0.84}{

\begin{tabular}{llcc}
\multirow{2}{*}{\textbf{Domain}} & \multirow{2}{*}{\textbf{Description}} & \multirow{2}{*}{\textbf{\begin{tabular}[c]{@{}c@{}}Range of \\ network sizes\end{tabular}}} & \multirow{2}{*}{\textbf{\begin{tabular}[c]{@{}c@{}}Number of \\ networks\end{tabular}}} \\
 &  &  &  \\ \hline
Friendship & \begin{tabular}[c]{@{}l@{}}Online friendship networks \\ (mostly Facebook) \end{tabular} & \begin{tabular}[c]{@{}c@{}}324-23,613 \\ (median: 4,065)\end{tabular} & 58 \\
Communication & \begin{tabular}[c]{@{}l@{}}Retweet, email and \\ reply networks\end{tabular} & \begin{tabular}[c]{@{}c@{}}96-33,696 \\ (median: 4,687)\end{tabular} & 43 \\
Collaboration & \begin{tabular}[c]{@{}l@{}}Co-authorship and collaboration\\ networks (mostly scientific)\end{tabular} & \begin{tabular}[c]{@{}c@{}}86-21,363 \\ (median: 3,621)\end{tabular} & 19
\end{tabular}
}
\label{table:networks}
\end{table}

The following topological metrics (for definitions see e.g.~\cite{barabasi2016network}) were calculated for each real network: assortativity, average clustering coefficient (avg\_clust), average degree (avg\_deg), average path length divided by the logarithm of the size (avg\_path\_log), density, global clustering coefficient (glob\_clust, four interval degree probabilities (idp\_41--idp\_44)~\cite{aliakbary2014quantification}, largest eigenvector centrality (max\_eigen), maximum degree (max\_deg), maximum edge and vertex betweenness centralities (max\_ebc and max\_vbc), number of edges and nodes and pseudo diameter divided by the logarithm of the size (p\_diam\_log). Note that a few of these metrics are highly influenced by the number of nodes, hence they have to be normalized. We follow the normalization procedure applied in~\cite{nagy2018data}. Obviously, there is a great deal of redundancy in the information provided by these graph metrics, thus based on the Spearman's correlations we narrow down the number of measurements to a smaller, non-redundant set of metrics (see Sec.~\ref{sec:structure}).

To investigate the descriptive ability of network models, we generate measurement-calibrated synthetic graphs for each real social network using four canonical network models: clustering Barab\'asi--Albert model (CBA)~\cite{holme2002growing}, stochastic block model (SBM)~\cite{peixoto2017nonparametric}, forest-fire model (FF)~\cite{leskovec2005graphs} and 2K model~\cite{gjoka2015construction}. The model calibration procedure can be formalized as follows. Let $G_T$ denote the real (target) network, $M(\theta)$ is a network model with parameter vector $\theta$, $G_{M(\theta)}$ is a realization of the model and $d$ is a distance function to compare graphs. We aim to adjust the parameters of the network model to minimize the distance, i.e.:
\begin{equation}
\label{eq:calibration}
    \theta^*  =  \argmin_\theta \, d(G_{M(\theta)},\, G_T). 
\end{equation}

To quantify the distance between two graphs, we calculate the Canberra distance of vectors of a reasonably chosen selection of graph metrics, i.e.  $d(G_1,G_2) = d_{\text{Can}}(f(G_1),f(G_2)),$ where $f = (f_1, f_2, \ldots, f_k)$ with $f_i$'s being real-valued functions defined on graphs corresponding to structural metrics and $k$ is the number of selected metrics.
The choice of Canberra distance is motivated by the work of Bonner \textit{et. al}~\cite{bonner2016efficient} and the metric selection is detailed in Sec.~\ref{sec:structure}. To minimize the distance, we use grid search optimization, for more details see~\cite{nagy2018data}. For the analyses and computations, we used the graph-tool~\cite{peixotographtool2014},
Networkx~\cite{hagberg2008exploring}, and igraph~\cite{igraph} Python modules.

\begin{figure}
    \centering
    \includegraphics[width=0.4935\linewidth]{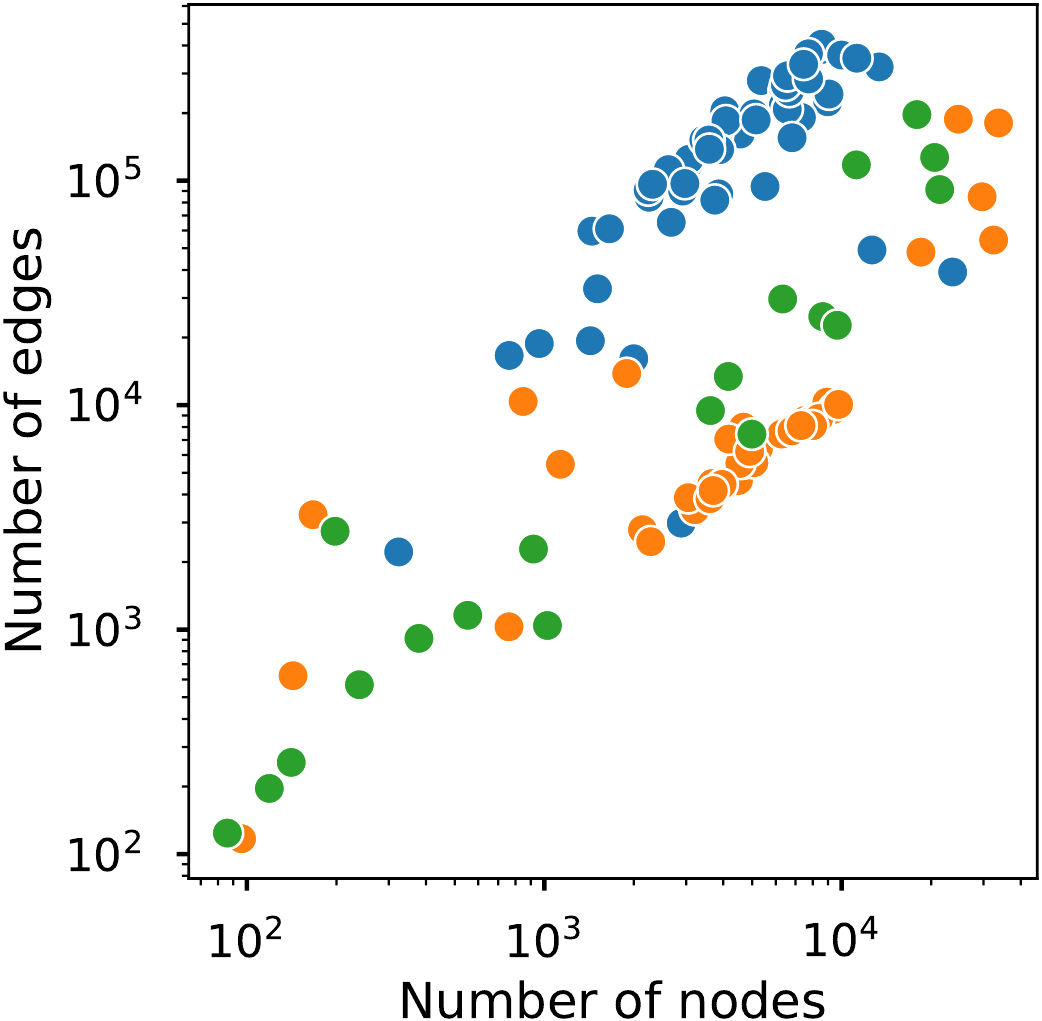}
    \includegraphics[width=0.4935\linewidth]{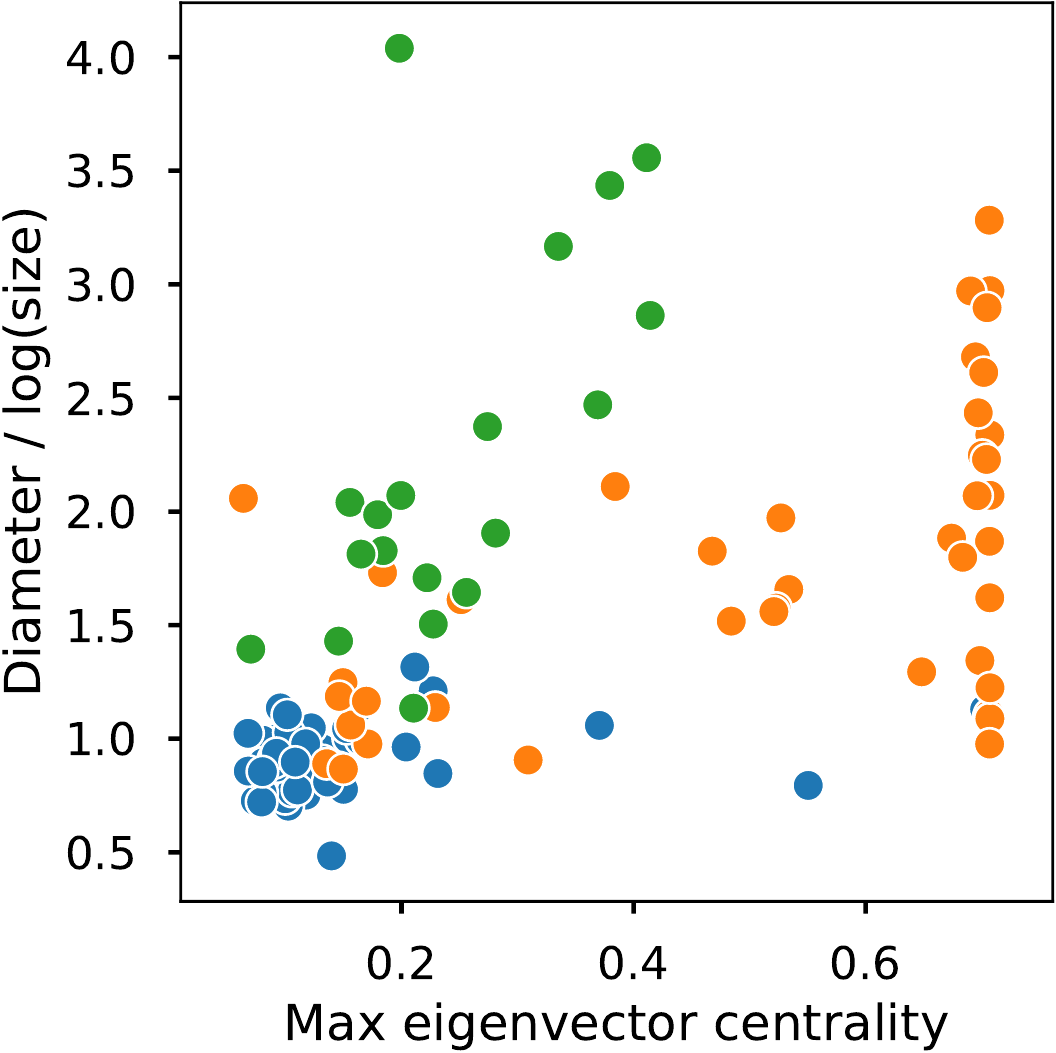}
    
    \vspace{0.4cm}
    
    \includegraphics[width=0.4935\linewidth]{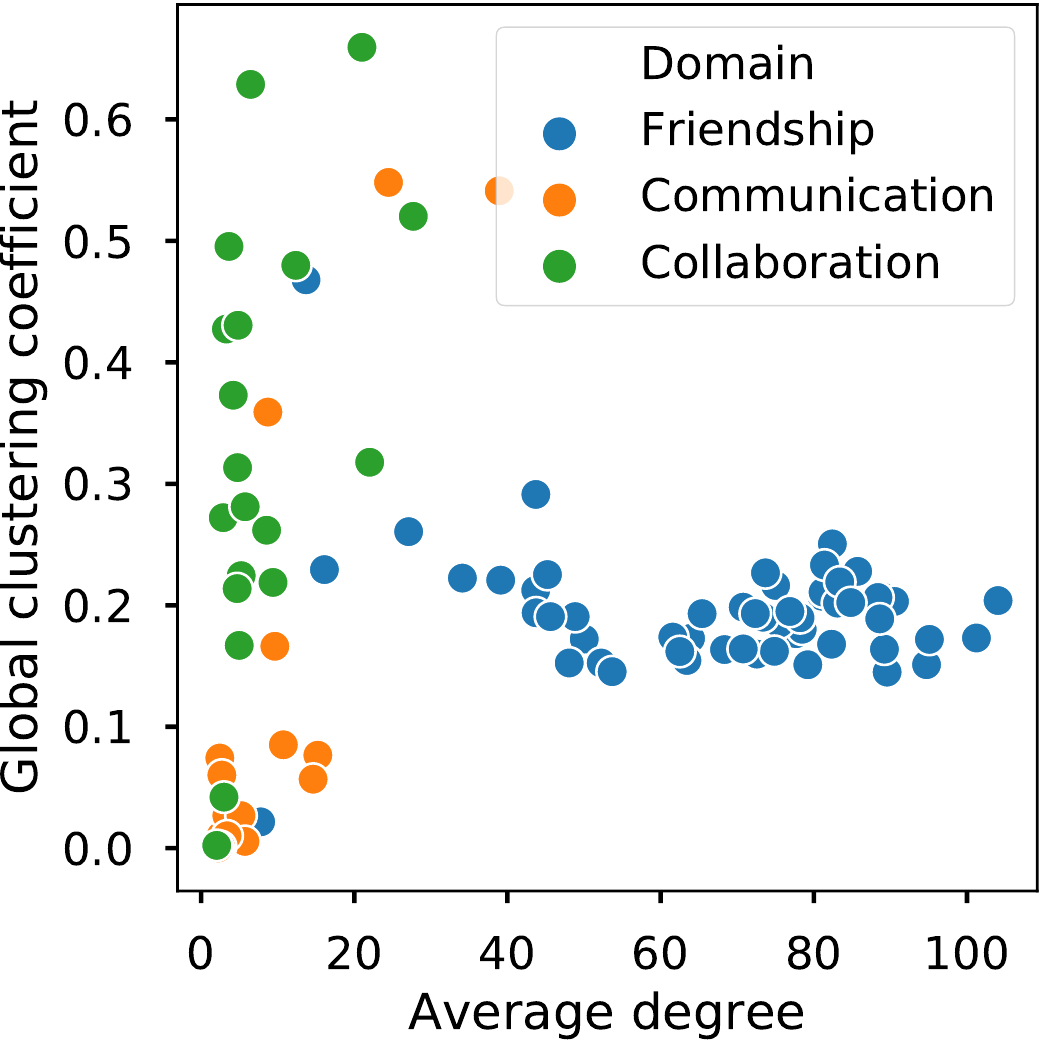}
    \includegraphics[width=0.4935\linewidth]{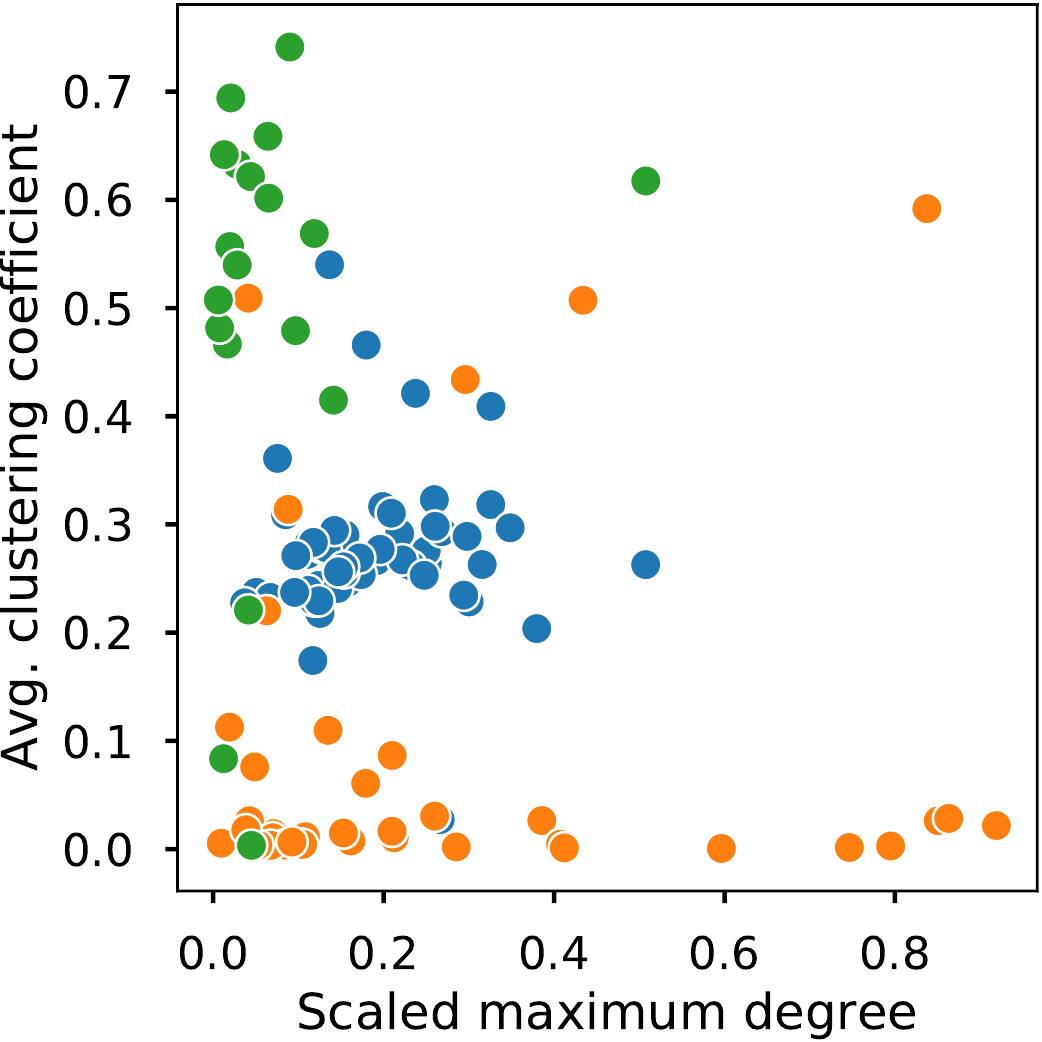}
    \caption{Scattering of the graph measurements of the real networks.}
    \label{fig:scatter}
\end{figure}

\section{Structural analysis of social networks}
\label{sec:structure}
In this section, we investigate the structural properties of social networks, particularly we study the (dis)similarity of the three domains: friendship, communication and collaboration networks. Furthermore, since this study is based on a large number of networks, we can also investigate the pair-wise relationships of the metrics across domains. To this end, we use Spearman's rank correlation, because it can measure non-linear relations, moreover, it is less sensitive to outliers. 

The scatter plots in Fig. \ref{fig:scatter} suggests that friendship (mostly Facebook) networks are densely connected, while communication (mostly Twitter) networks are extremely sparse. Moreover, collaboration and communication networks can be easily distinguished by the clustering coefficients. The fact that the retweet networks are often star-shaped -- as also illustrated in ~\cite{chen2018nature} -- is clearly observable in the scattering of their maximum eigenvector centralities and normalized maximum degrees.    

\begin{figure}
    \centering
    \includegraphics[width=0.72\linewidth]{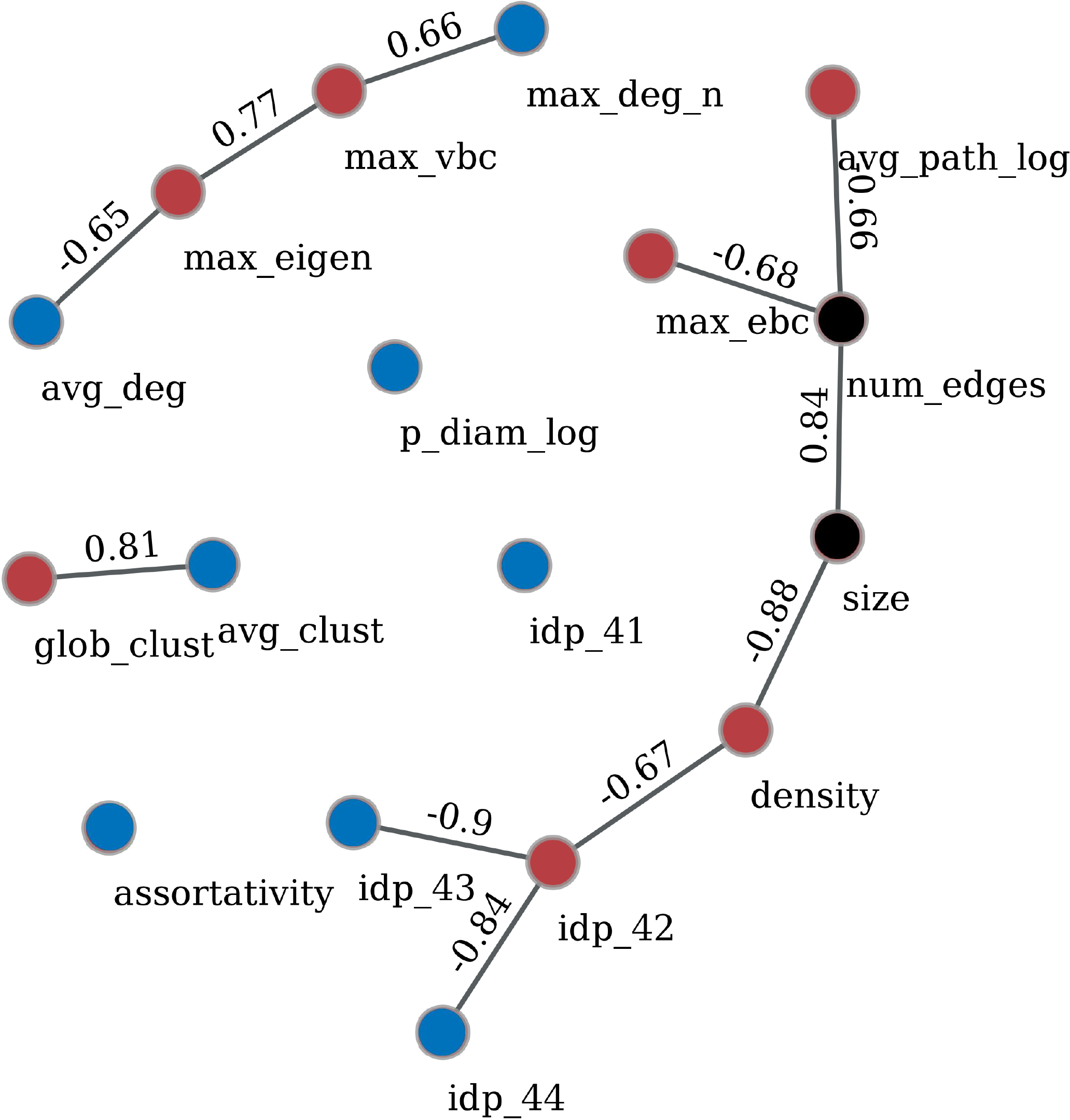}
    \caption{The correlation network of structural metrics. Two nodes are connected if the domain-averaged absolute Spearman's rank correlation of the corresponding metrics is above $0.65$. The exact values of the correlations are written on the edges. The blue nodes form the set of selected metrics, the black nodes and their neighbors are excluded in the selection procedure.}
    \label{fig:corr_graph}
\end{figure}

Fig. \ref{fig:corr_graph} shows the correlation network of the metrics. Based on the correlation network, we aim to select a non-redundant subset of metrics with high descriptive ability. We select the metrics such that the resulting set is a maximal independent set in the network that is obtained from the correlation network (Fig.~\ref{fig:corr_graph}) by excluding the nodes that are strongly correlated with the number of nodes and edges. We believe that size (and the metrics that are highly correlated with it) should not influence the similarity of two networks, i.e. just because two networks have the same number of nodes it does not mean that they are similar, hence we aim to have a set of size-independent metrics. Our selection of metrics is listed in Table~\ref{selected_attributes}.

% the high correlation with the size implies significant trivial distinguishing power with respect to network domains due to the density difference of typical networks from different domains. 

\begin{table}[]
\centering
\caption{The selected structural metrics and the nominal variables indicating the origin of the networks.}
% \scalebox{0.8}{
\begin{tabular}{ll}
\textbf{Name} & \textbf{Description} \\ \hline
assortativity & Assortativity coefficient \\
avg\_clust & Average local clustering coefficient \\
avg\_deg & Average degree \\
p\_diam\_log & \multirow{2}{*}{\begin{tabular}[c]{@{}l@{}}Pseudo diameter divided by the \\ logarithm of the size\end{tabular}} \\
 &  \\
idp's (1,3,4) & Interval degree probabilities \\
max\_deg\_n & Maximum degree divided by the size \\\cline{2-2} 
domain & \multirow{2}{*}{\begin{tabular}[c]{@{}l@{}}Domain of the real networks:\\ friendship, communication, collaboration \end{tabular}} \\
 &  \\
category & \multirow{4}{*}{\begin{tabular}[c]{@{}l@{}}Indicates whether the graph is real or model generated,\\ together with the type of the model: \\  real (original), 2K, CBA (clustering Barab\'asi--Albert), \\FF (forest-fire), SBM (stochastic block model)\end{tabular}} \\
 & \\
 & \\
 
\end{tabular}
% }

\label{selected_attributes}
\end{table}

 From the domain-specific correlation heatmaps (due to their large size only available in the supplementary material~\cite{own_data2})  we can conclude that the correlation profiles vary across domains. However, there are some universal relations that can be observed in Fig.~\ref{fig:corr_graph}. For example, the maximum centrality related measures are highly correlated, the interval degree probabilities are also connected, while it is surprising that the normalized average path length and diameter are not so correlated. 

% the interval degree probabilities are highly correlated in each domain. We can observe that size, density and clustering related metrics have strong correlations for friendship networks. For communication networks, there is a large set of highly correlated metrics (indicated by the block diagonal structure of correlation matrix) including assortativity, clustering coefficients, interval degree probabilities, maximum of node centralities (degree, eigenvector and betweenness centrality). The most correlated metrics for collaboration networks are the average degree, average path length, diameter, and maximum betweenness centralities.

% \begin{figure}
%     \centering
%     \includegraphics[scale=0.52]{Scatterplots.pdf}
%     \caption{On the left: The scatter plot of the networks from the food web domain with respect to the maximum eigenvector centrality and the assortativity metrics. Both the original (real) and the model-generated networks are illustrated (see the colours). On the right: The projection of the real networks to the first two principal components show a clear distinction of the different domains.}
%     \label{fig:pca}
%     % \vspace{-10pt}
% \end{figure}

\begin{figure*}
    \centering
    \includegraphics[width=0.32\linewidth]{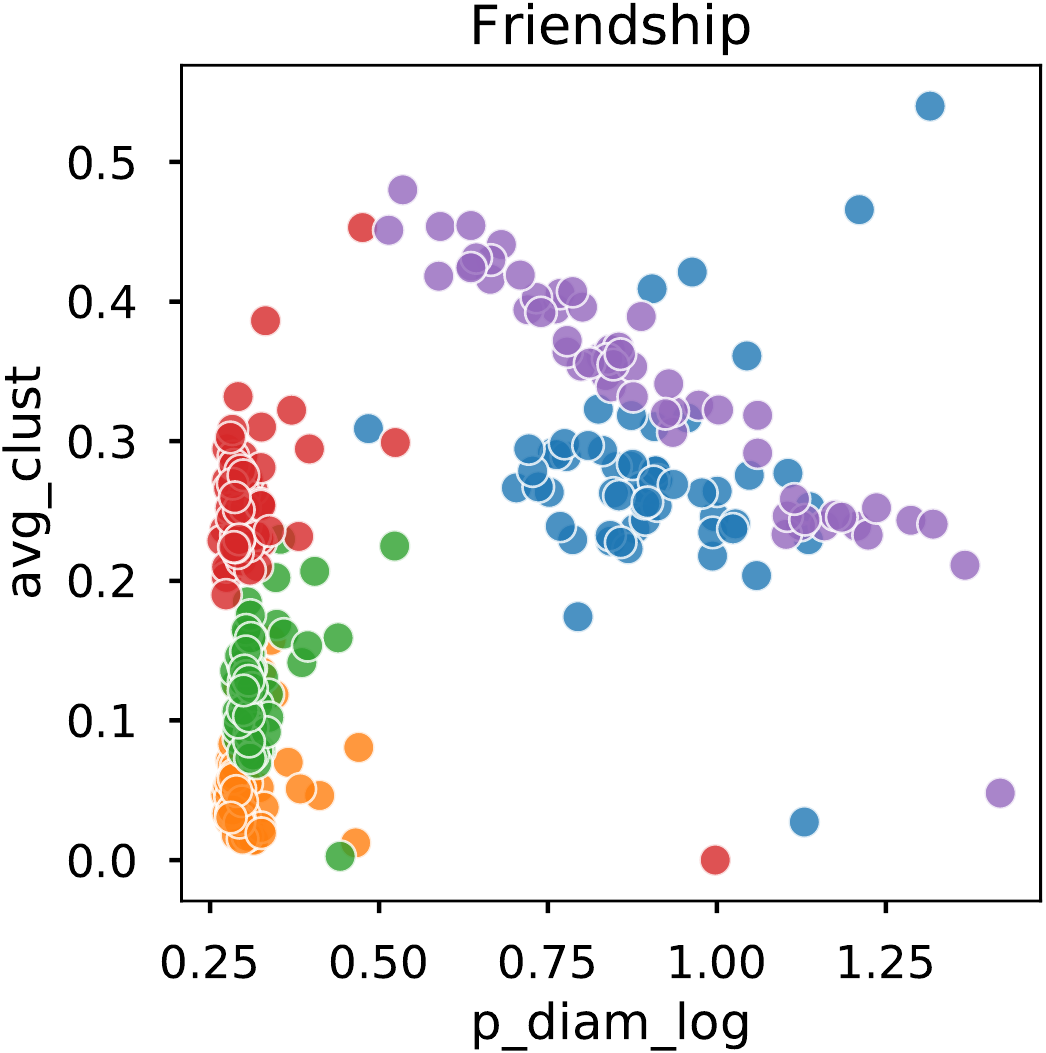}
    \includegraphics[width=0.32\linewidth]{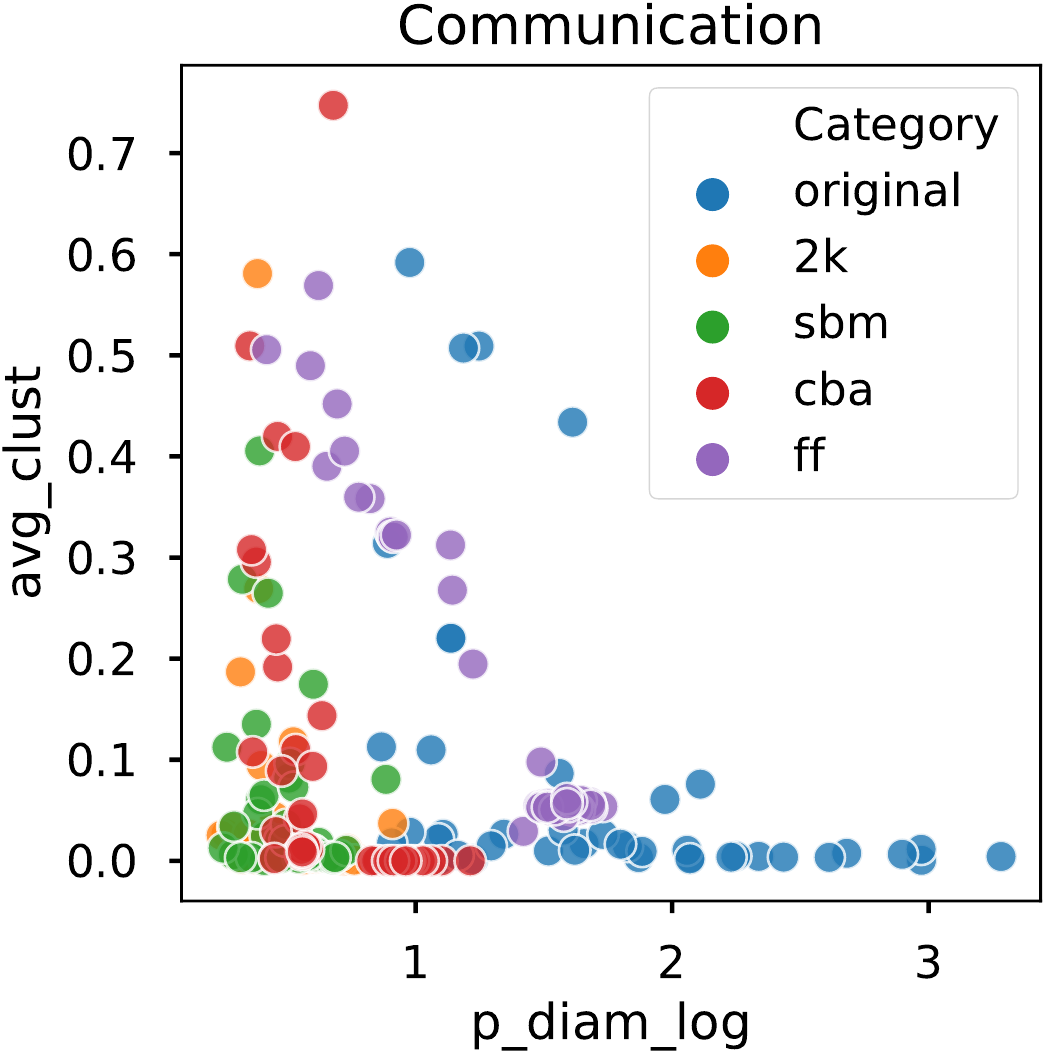}
    \includegraphics[width=0.32\linewidth]{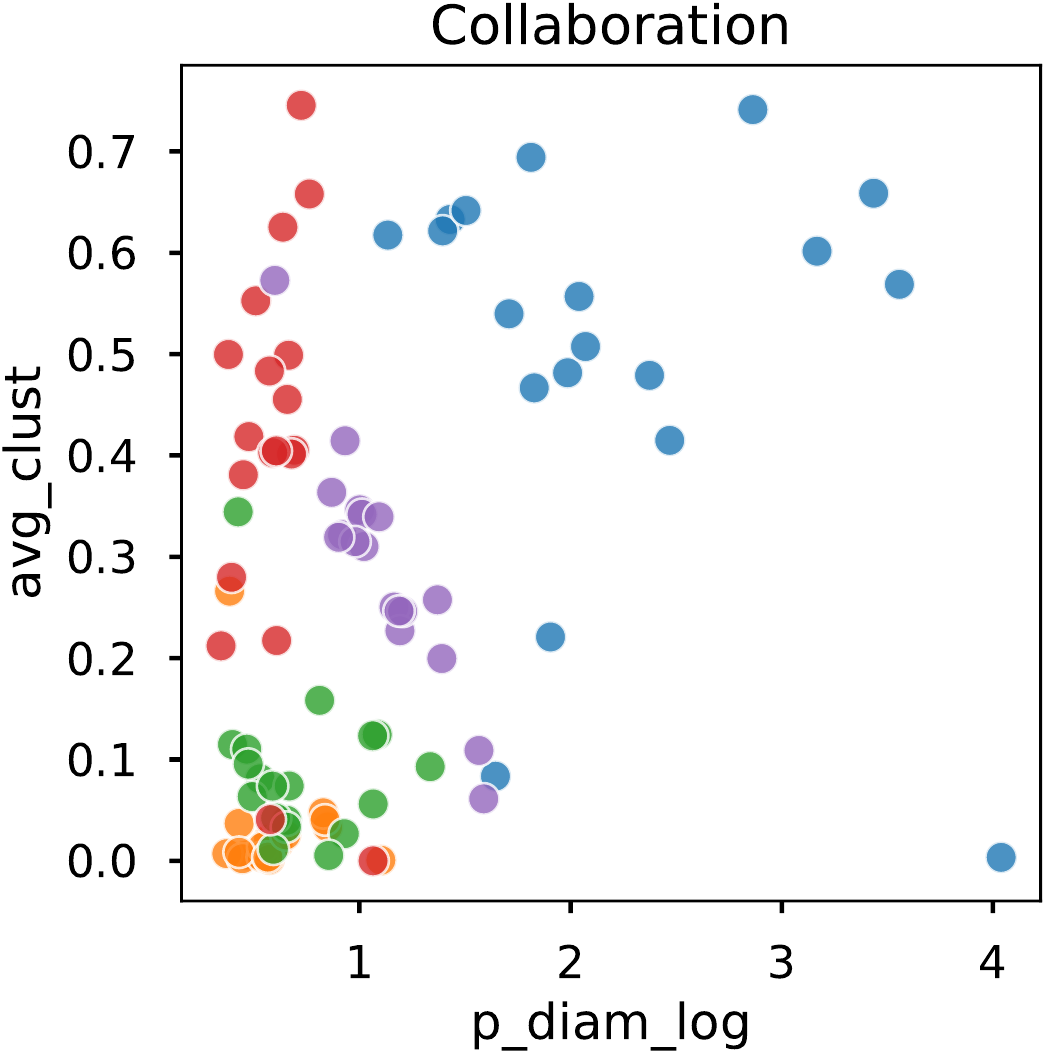}
    \caption{The structural properties of real networks that the models cannot capture efficiently.}
    \label{fig:weakpoints}
\end{figure*}
\begin{figure*}
    \centering
    \includegraphics[width=0.32\linewidth]{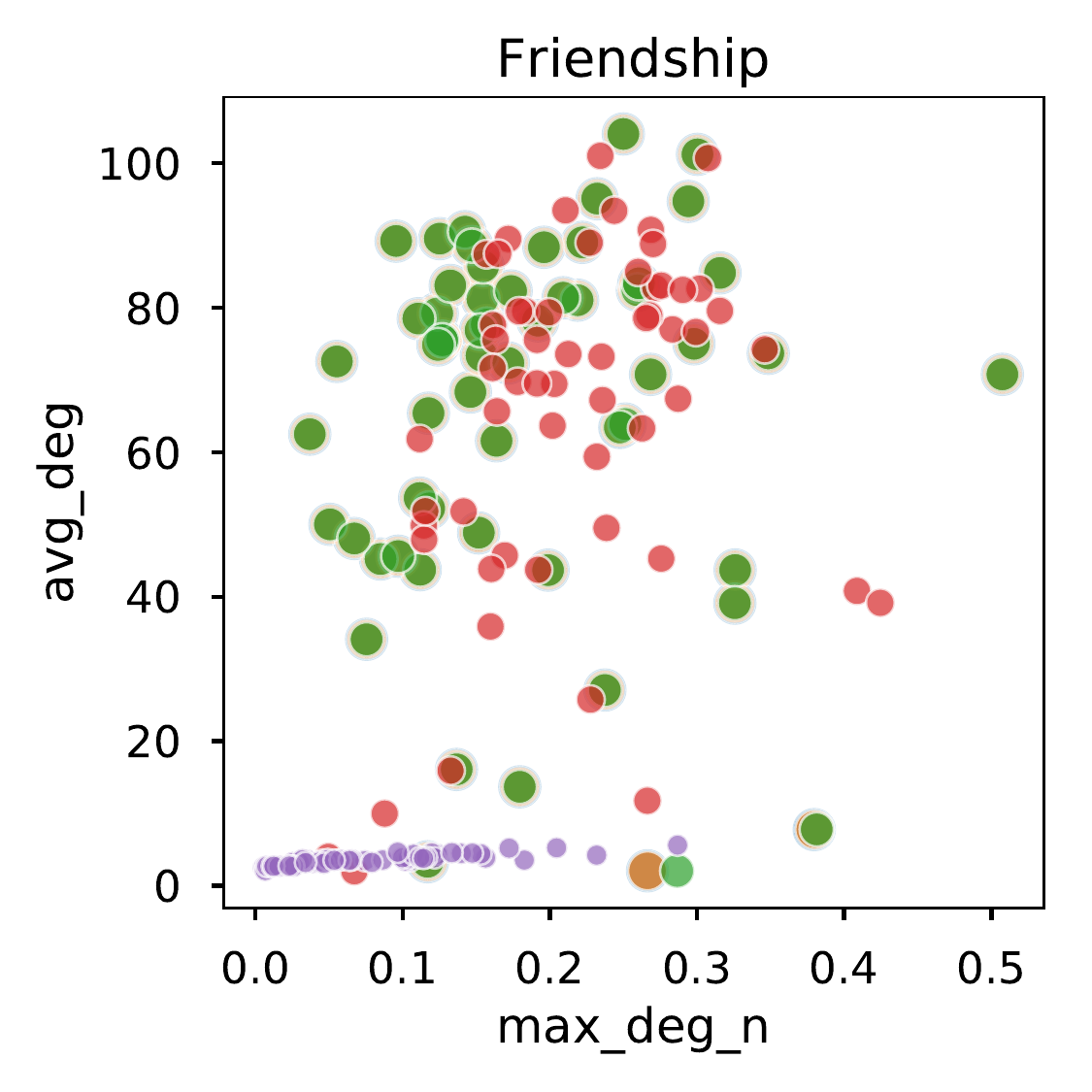}
    \includegraphics[width=0.32\linewidth]{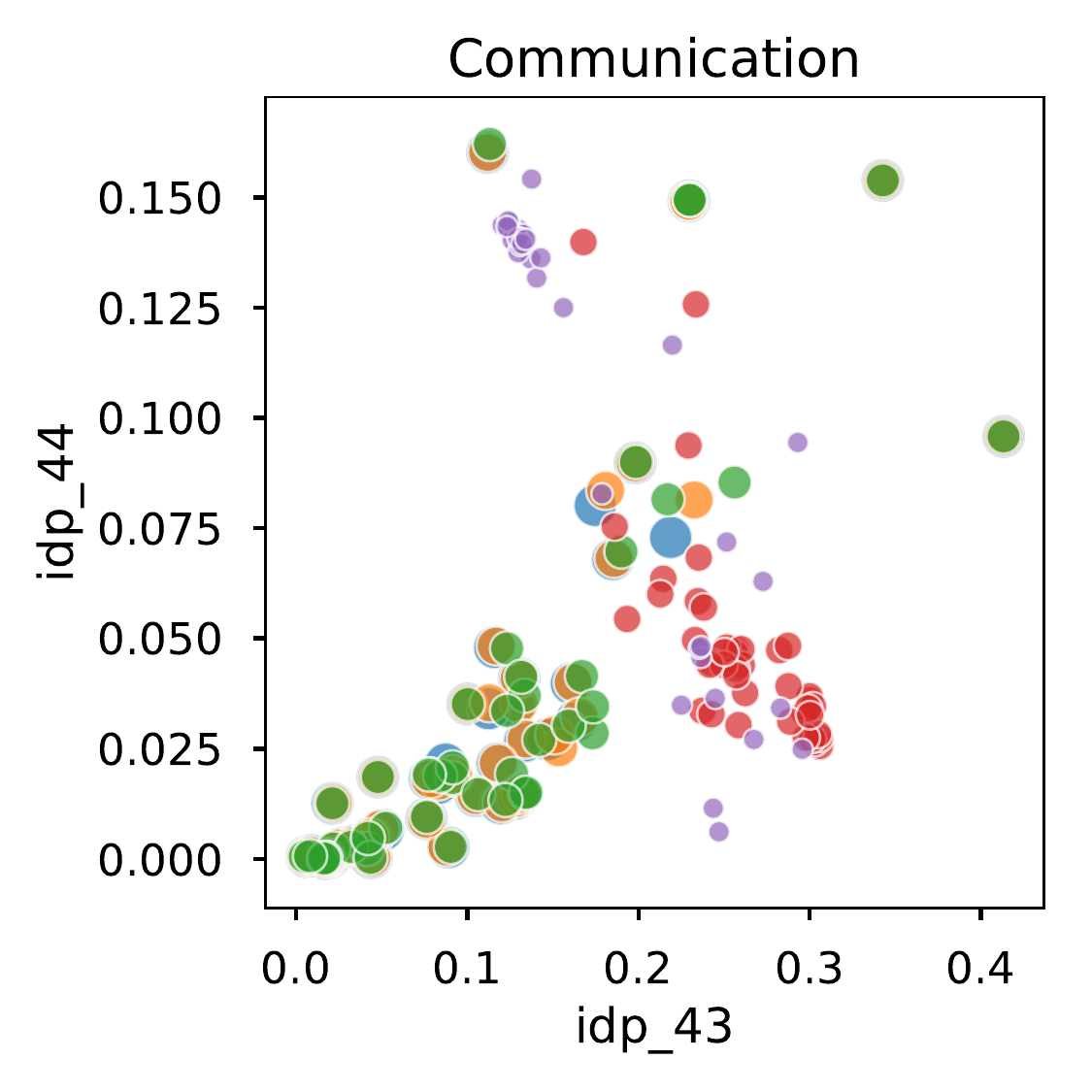}
    \includegraphics[width=0.32\linewidth]{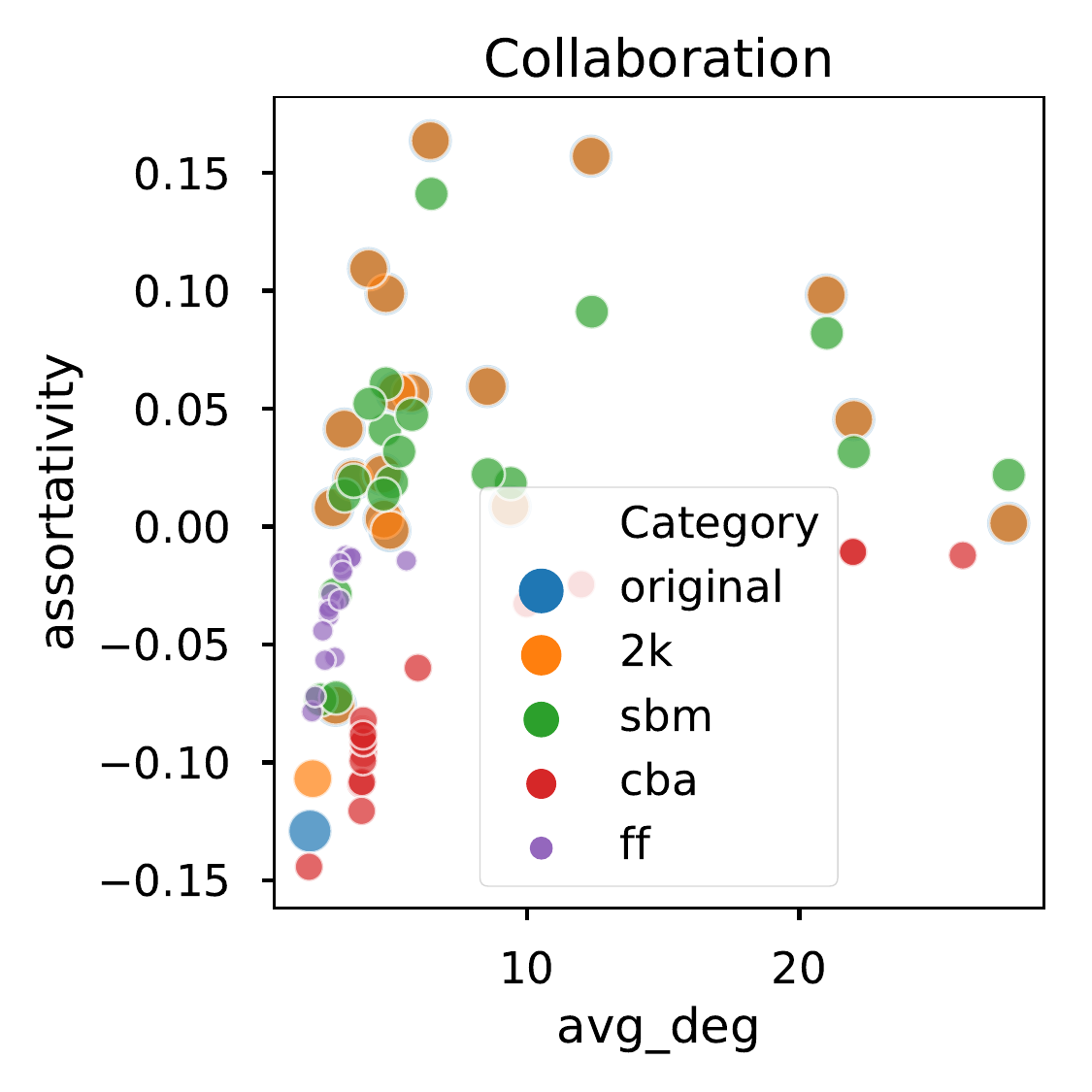}
    \caption{The structural properties of real networks that models can capture accurately. Different sized dots are used  only to be able to see the overlaps more clearly.}
    \label{fig:strengths}
\end{figure*}

\section{Measurement-calibrated synthetic networks}
\label{sec:synthetic}
In this section, we study how realistically we can synthesize social networks, particularly we aim to identify the characteristics of real social networks that the four selected network models can or cannot capture. Note that in~\cite{nagy2018data} we have shown that the network models are stable enough to perform measurement-based calibration with respect to a given target social network as formulated in~\eqref{eq:calibration}.

Our two-stage approach can be summarized as follows: first, we fit each model to each real network and generate a synthetic graph with the calibrated parameters, i.e. each real network has four model-generated counterparts. Secondly, after we calculated the graph metrics detailed in Table~\ref{selected_attributes} of the newly generated graphs, we compare the synthetic and real networks through their graph metrics.

% \begin{figure}
% %\vspace{-13pt}
%     \centering
%     \includegraphics[scale=0.61]{importantvars.pdf}
%     %\vspace{-5pt}
%     \caption{Scatter plot of real and modeled graphs along the most important variables with respect to the category prediction problem determined by the machine learning algorithms in each domain.}
%     \label{fig:importantvars}
%     %\vspace{-15pt}
% \end{figure}

% \begin{figure*}
%     \centering
%     \includegraphics[height=0.45\textheight]{boxp.pdf}
%     \caption{Boxplots of the metrics across different models. The modelled graph is a social network with 17903 nodes. Since the 2K model captures all the information about the joint degree distribution, it is capable of exactly fitting the assortativity, density and the skewness of the degree distribution. Overall the models could not estimate the average clustering coefficient well, except for the WS model that was fitted according to this metric. Clearly, the density of the models can be directly adjusted, except for the Duplication Divergence model. The skewness of the degree distribution of the CBA and WS models is around zero since these models are not scale-free. Generally, the DD and CBA models are the least stable.}
%     \label{fig:boxplotsSocial}
% \end{figure*}

In order to compare how realistic the models are and how modelable the domains are, we calculated the mean Canberra distance between the numeric attributes of the original and the model-generated graphs for each domain separately. Note that in this case we also take into consideration the size of the original and the generated networks, since in contrast to the other applied models, the SBM does not necessarily generate connected graphs, hence we extracted the largest connected component, which may be smaller than the target real network. 

\begin{figure*}[!htb]
  \includegraphics[width=0.32\linewidth]{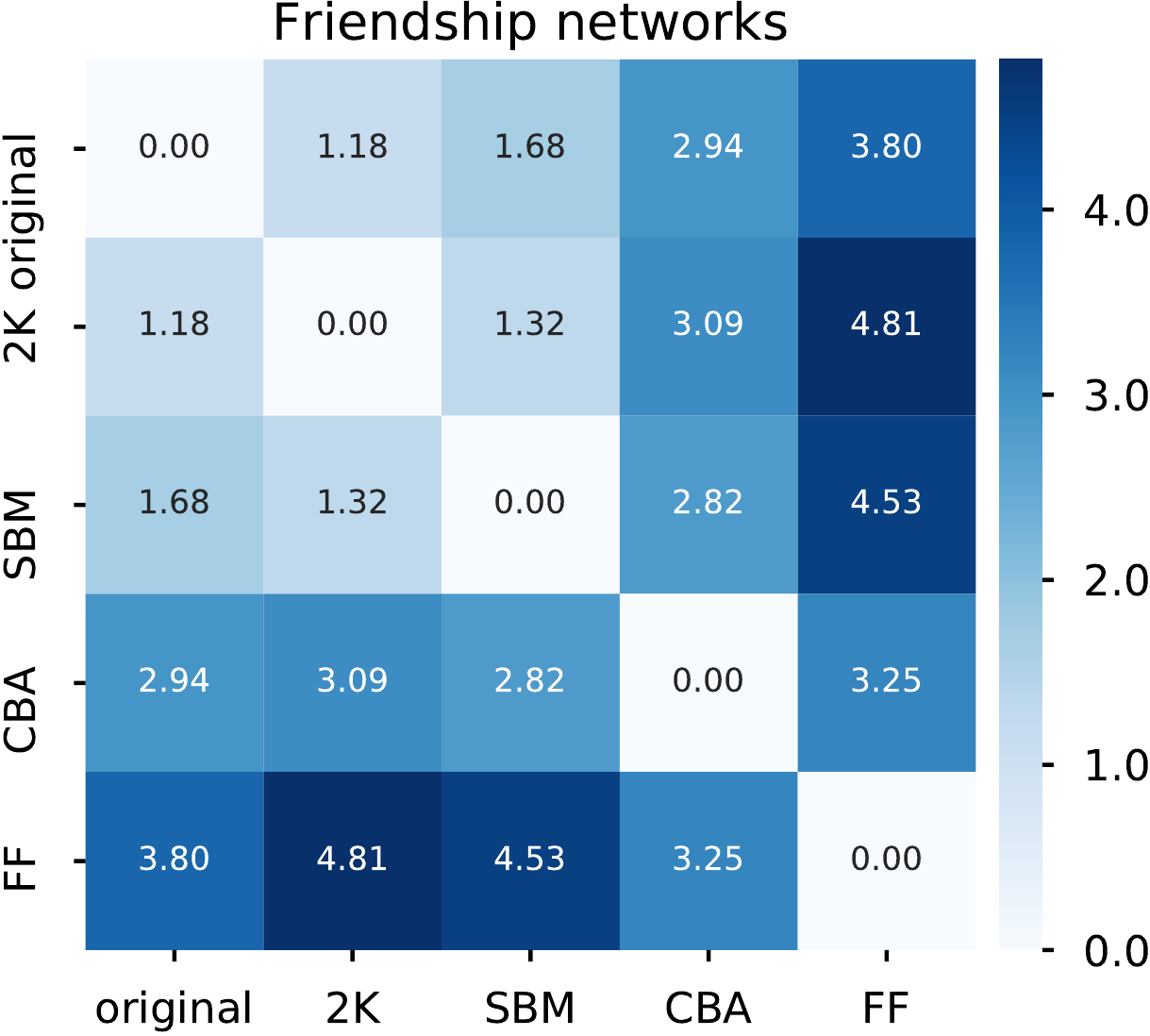}
  \includegraphics[width=0.32\linewidth]{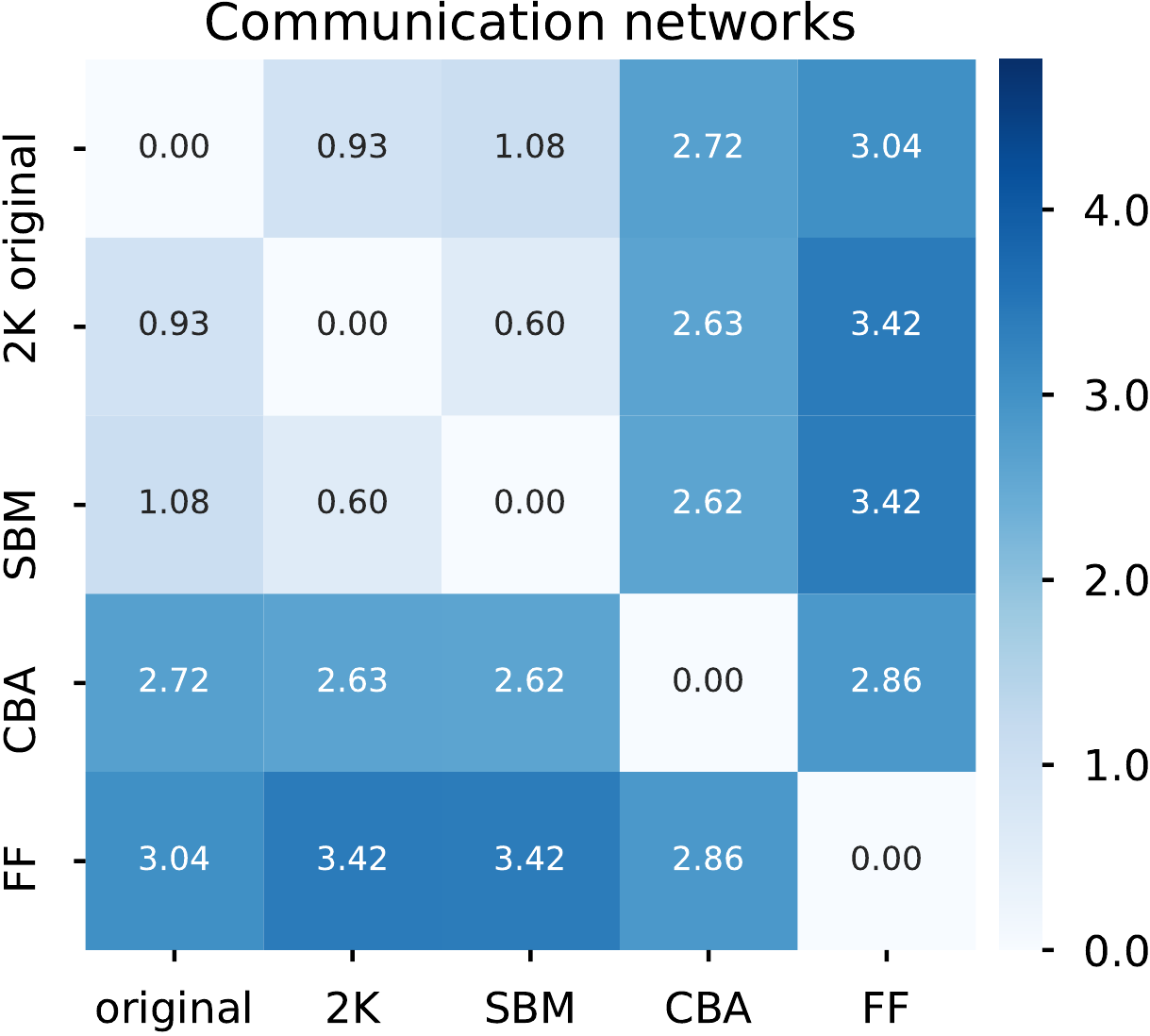}
  \includegraphics[width=0.32\linewidth]{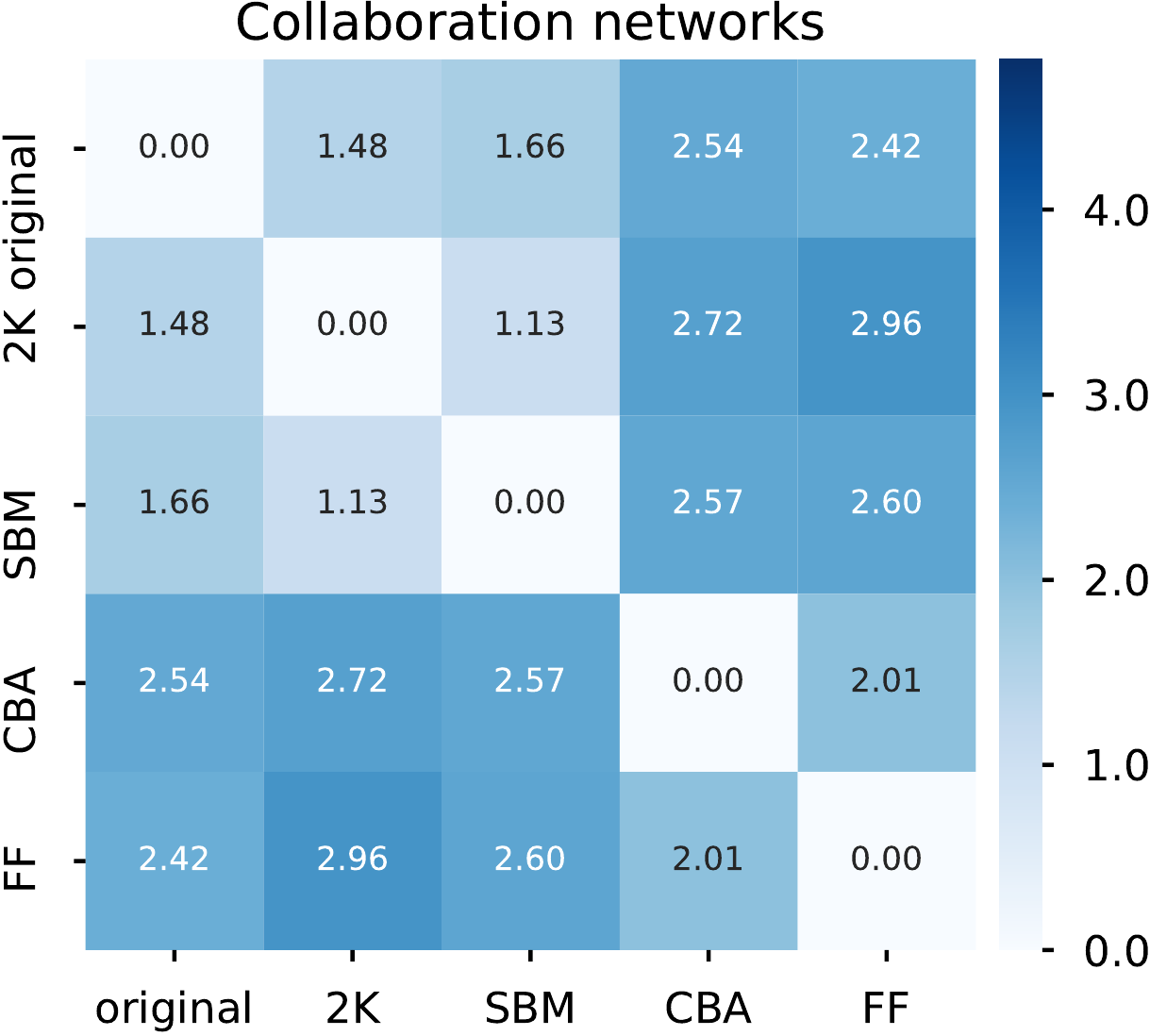}
\caption{The mean Canberra distances between the original and the model-generated graphs across domains.}
\label{fig:canberra}
\end{figure*}

The results of the domain-averaged Canberra distances are shown in Fig.~\ref{fig:canberra}. It can be seen that on every domain the SBM and 2K model efficiently capture the structural properties of real networks. Furthermore, on average, communication networks are the easiest and friendship networks are the most difficult to mimic, however, the lowest accuracy of the two best performing models were achieved on  collaboration networks. It is also important to mention that independently of the domains, SBM and 2K model generate similar graphs, which is also clearly visualized in Fig.~\ref{fig:weakpoints} and Fig.~\ref{fig:strengths}. Moreover, from Fig.~\ref{fig:weakpoints} we can also conclude that none of the models can capture the relationship of the average clustering coefficient and normalized diameter. While the forest-fire model approximated the values of these two metrics the most accurately, it failed to capture other pair-wise connections such as the maximum degree and the average degree. It is important to note, that in general, models could mimic the structural properties of real networks, especially SBM and 2K model as it can be seen in Fig.~\ref{fig:strengths}. What is more, these two models can capture the exact values of degree distribution related metrics by their generation mechanisms, thus the dots corresponding to the original, 2K, and SBM graphs overlap each other. 
% \begin{figure*}
% %\vspace{-2pt}
%     \centering
%     \includegraphics[scale=0.398]{distheatmap.pdf}
%     %\vspace{-15pt}
%     \caption[caption]{The average Canberra distance between the original and the model-generated graphs. In every domain, the 2K model efficiently captured the features of the original networks, the second-best performing models are varying across domains though....
%   The figure also shows that food webs are the easiest to model: 2K, CBA and DD algorithms generated graphs are very similar to the real ones.}a
%     \label{fig:distheatmap}
%     %\vspace{-15pt}
% \end{figure*}

\section{Conclusion}
\label{sec:conc}

In this paper, we studied 120 large social networks and their measurement-calibrated model-generated counterparts to identify the structural properties that discriminate either the subdomains of social networks or the synthetic graphs from real networks. We illustrated what topological properties of real social networks the models can or cannot capture, and we found that models - except for the forest-fire - unable to generate graphs with a relatively high diameter and high clustering coefficient simultaneously. However, overall the 2K and the degree corrected stochastic block models can efficiently mimic the social networks.

A natural extension of this work is to compare the descriptive ability of graph measurements to state-of-the-art graph embedding methods such as Graph2Vec~\cite{narayanan2017graph2vec}.

\section*{Acknowledgment}

The research reported in this paper was supported by the BME - Artificial Intelligence FIKP grant of EMMI (BME FIKP-MI/SC). The publication is also supported by the EFOP-3.6.2-16-2017-00015 project entitled "Deepening the activities of  HU-MATHS-IN, the Hungarian Service Network for Mathematics in Industry and Innovations". The research of R. Molontay was partially supported by NKFIH K123782 research grant.

\bibliographystyle{IEEEtran}
\bibliography{ref.bib}
\end{document}